\documentclass[12pt]{article}

\usepackage{times}
\usepackage{epsfig}
 
\begin{document}

\begin{center}
{\large{A New Model of Solar Neutrinos in Manifest Violation of CPT Invariance}}

{\small{}R. S. Raghavan}

{Bell Laboratories, Lucent Technologies, Murray Hill NJ 07974 USA}

{INFN Laboratori Nazionali del Gran Sasso, Italy}

\today
\end{center}

\begin{abstract}
The large mixing (mass) (LMA) -MSW model of solar neutrinos ($\nu _{\mathrm{e}}${) is now widely held to be near definitive, based on global consistency with data.
No physical effect, however, compels its }\textit{uniqueness}{.
The present search for an explicitly testable competitive model was stimulated by a surprising finding-- the high energy part of the standard solar model (SSM)}
$^{\mathrm{8}}$B $\nu_{\mathrm{e}}${ spectrum can be scaled very precisely to observed flux levels }\textit{without measurable shape distortion} {via sensitive combinations of long wavelength flavor conversion in vacuum and a }
$^{\mathrm{{8}}}${B }$\nu_{\mathrm{e}}$ { flux }
$\phi {(\mathrm{B}) < } \phi${(B: SSM).
Pursuantly, a new ``astroparticle'' model
with the relatively specific parameters }
$\Delta {m} ^{\mathrm{2}} {=76-78 } \mu${eV}$^{\mathrm{2}} (10^{\mathrm{-12}}${ eV}$^{\mathrm{2}}${);
sin}$^{\mathrm{2}} 2\theta =$ 0.58--0.56; $\phi${(B) = 0.83--0.81}$\phi (^{\mathrm{8}}${B:SSM)}
{coupled with modest changes in the SSM, offers a viable solution consistent with data.
Because KamLAND has set }$\Delta${m}$^2 \sim 50 \times 10^6 \mu${eV}$^2$ {,
sin}$^{2} 2 \theta \sim 1$ for \textit{anti}\textrm{neutrinos, our model manifestly violates CPT invariance.
The model predicts new distortional effects in solar neutrino spectra via }$\nu_{\mathrm{{e}}}${-e scattering signals in the window 3--5 MeV,
even though the spectrum is flat above 5 MeV\@.
This window is
accessible to experiment for the first time in KamLAND\@.
New experiments are proposed to observe the more dramatic charged-current spectral effects. }
\end{abstract}

\section{Introduction}

{Experimental data on solar neutrinos (}{\(\nu\)}$_{\mathrm{{e}}}${)}$_{\mathrm{{ }}}${ has accumulated rapidly in the last few years by the operation of several large detector facilities such as Super-Kamiokande (SK), the Sudbury Neutrino Observatory (SNO) and KamLAND\@.
The data from these detectors were precise and informative.
The neutral current signals  (NC) from }$^{\mathrm{{8}}}${B neutrinos in SNO
\cite{ahmad}
showed that }{\(\nu\)}$_{\mathrm{{e}}}${ flavor was not conserved because the flavor-blind NC rate explicitly exceeded the charged current (CC) }{\(\nu\)}$_{\mathrm{{e}}}${ signal rate.
The absolute NC rate was close to the }$^{\mathrm{{8}}}${B }{\(\nu\)}$_{\mathrm{{e}}}${ flux of the Standard Solar Model (SSM)\cite{bahcall1}.
The }($\nu${-electron) scattering (ES) signal in SK \cite{fukuda} indicated the }\textit{{type}}{ of flavor conversion with a highly precise energy-independent ES spectrum in the measured window 5-15 MeV\@.
This result severely restricted the possible }{\(\nu\)}$_{\mathrm{{e}}}${ models.
Global analyses of solar }{\(\nu\)}$_{\mathrm{{e}}}${ data pinpointed matter conversion (MSW effect) in the large mixing (mass) (LMA-MSW) regime as the most favored model of solar neutrinos\cite{pakvasa_valle}.
Finally, KamLAND observed oscillations of }\textit{{terrestrial}}{ }\textit{{anti}}{neutrinos\cite{eguchi}.
The resulting parameters were approximately consistent with LMA-MSW neutrinos, as expected by the invariance of CPT symmetry.
The solar neutrino problem therefore appears resolved.}

{In view of the preponderance of evidence of global consistency for the LMA-MSW model, why search for an alternate model? In principle, however, no physical effect specific to LMA has yet been observed to prove its}\textit{{ uniqueness. }}{ The NC/CC result from SNO\cite{ahmad}, shows only that the }{\(\nu\)}$_{\mathrm{{e}}}${'s oscillate but it does not specify the mechanism as LMA-MSW. The lack of a uniqueness proof becomes a central issue (not just a point of principle) if an }\textit{{explicitly testable }}{competitive model can be invented.}

{In this work I show that such a solar neutrino model can be constructed. The uniqueness of }\textit{{this }}{model can be proved or disproved by a variety of effects at signal energies $<$5 MeV\@. In particular, in contrast to LMA-MSW, sizable shape distortions are predicted in the 3--5 MeV window in ES spectra even though the shape is still flat above 5 MeV\@.
Such an effect in solar neutrino spectra is predicted for the first time.
The 3--5 MeV window is now experimentally observable for the first time in a liquid scintillation device such as the already operational KamLAND detector; thus, the above prediction can be readily subjected to experimental test. }

{The physical mechanism of the present model is long wavelength oscillations (LWO) in vacuum, discovered by a new approach based on ``spectral engineering''. Although the underlying parameters were previously detected in the general dragnet of global analysis programs \cite{fogli,krastev_petcov} the }\textit{experimental effects} of the model are new, unsuspected and distinctly different from those of generic vacuum oscillations.

The $\nu_{\mathrm{{e}}}${ parameters of the model, very different from LMA, imply manifest violation of CPT invariance in view of the KamLAND results.  Whether this is a fatal flaw is at present unclear from the standpoint of basic theory. The suggested experimental tests of the new model, however, offer a decisive resolution of the question, thus they are doubly  important.}

{The present work has its origins in a question posed in 1994, if it was possible to conceive a neutrino oscillation model elusive enough to }\textit{{escape}}{ detection in experiments at the time.
The answer was yes, if solar neutrinos oscillated at only the lowest energies.
In very long wavelength (VLWO) oscillations in vacuum with }$\Delta$m$^2 < 10 \mu$eV$^2 (10^{-12}$eV$^2$), $^8$B neutrinos are hardly converted;
neutrinos $<$1 MeV are converted, and explain the $^{71}${Ga results\cite{raghavan1}.
Thus, the }$^{\mathrm{{8}}}${B flux deficit  observed in the }$^{\mathrm{{37}}}${Cl}{ data can only be produced by an initial flux }\textit{{reduced astrophysically to $\sim$1/3.}}
{ This scenario, called ``Just-so 2''\cite{bahcall2}, was the first model to be ruled out \cite{barger} by the SNO NC result -- the initial flux is apparently the full SSM flux, thus, }\textit{conversion }{is centrally responsible for the flux deficit.
Such an effect is available via LMA-MSW especially since it also supplies the crucial undistorted SK spectrum.
Thus, conversion models at two extremes produce the observed undistorted }$^{\mathrm{{8}}}${B spectrum in principle, via: 1) VLWO with no }$^{\mathrm{{8}}}${B }{\(\nu\)}$_{\mathrm{{e}}}${ flavor conversion but a flux reduction by a factor $\sim$0.36; and 2) LMA-MSW with }{\(\nu\)}$_{\mathrm{{e}}}${ survival of $\sim$0.36 but the full  initial SSM }$^8${B flux. }

{The question then arises whether a scenario could be engineered with a }\textit{{mixture}}{ of }\textit{{both}}{ flavor conversion }\textit{{and}}{ a smaller initial flux that leaves the high energy part of the }$^{\mathrm{{8}}}${B }{\(\nu\)}$_{\mathrm{{e }}}${ spectrum }\textit{{shape}}{ }\textit{{invariant}}{ so that all the experimental consequences (at $>$5 MeV)  follow automatically. By definition, this does need a smaller initial }$^{\mathrm{{8}}}${B flux than the SSM/SNO value, however, in the spirit of global analyses made so far, one could tolerate in principle a lower flux up to the 2}{\(\sigma\)}{ lower limit of the SNO NC result, i.e., $>$ 75\% of the SSM flux. }

{The surprising initial discovery was that a shape invariant high energy }
$^8${B neutrino spectrum could indeed be engineered to a high degree of precision (within a few percent) by a sensitive choice of the initial flux and mass-mixing parameters in a very narrow band }$\Delta${m}$^2${= 70-80 }{$\mu$}{eV}$^2${, $\sim 10$ times larger than the
VLOW-just-so2 regime \cite{raghavan1}.
The oscillations in the new regime are thus termed long-wavelength or LWO in this paper.
The new mass-mixing band is very different from LMA-MSW regime. In view of the KamLAND result, one thus immediately faces the problem of violation of CPT invariance by neutrinos. Nevertheless, work proceeded undeterred. Recent theoretical ideas appear to allow that the otherwise rigid invariance of CPT may be relaxed in principle in the neutrino sector\cite{barenboim}.}

\section{Spectral Engineering}

{Fig.1 illustrates the central thesis of this work. It shows a series of }$^{8}${B }{$\nu_\mathrm{e}$ spectral shapes engineered via LWO to reproduce at high energies, the rate and shape of the LMA-MSW (thick red) solution as closely as possible. The LMA shape is defined as the standard SSM shape (thin blue) scaled by the factor 0.365 (0.349(22), the CC flux at SNO).
The convenient definition ``LMA = 0.365 SSM'' agrees very closely in the
$\phi$($^8$B) and the flat spectral shape observed above 5 MeV\@.
The other spectra are derived with ``shape invariant'' LWO parameter combinations (see box and caption of Fig.1) chosen individually and interactively on-line. }

\begin{figure}[ht]
%\vspace*{-1.5in}
\centerline{\epsfig{file=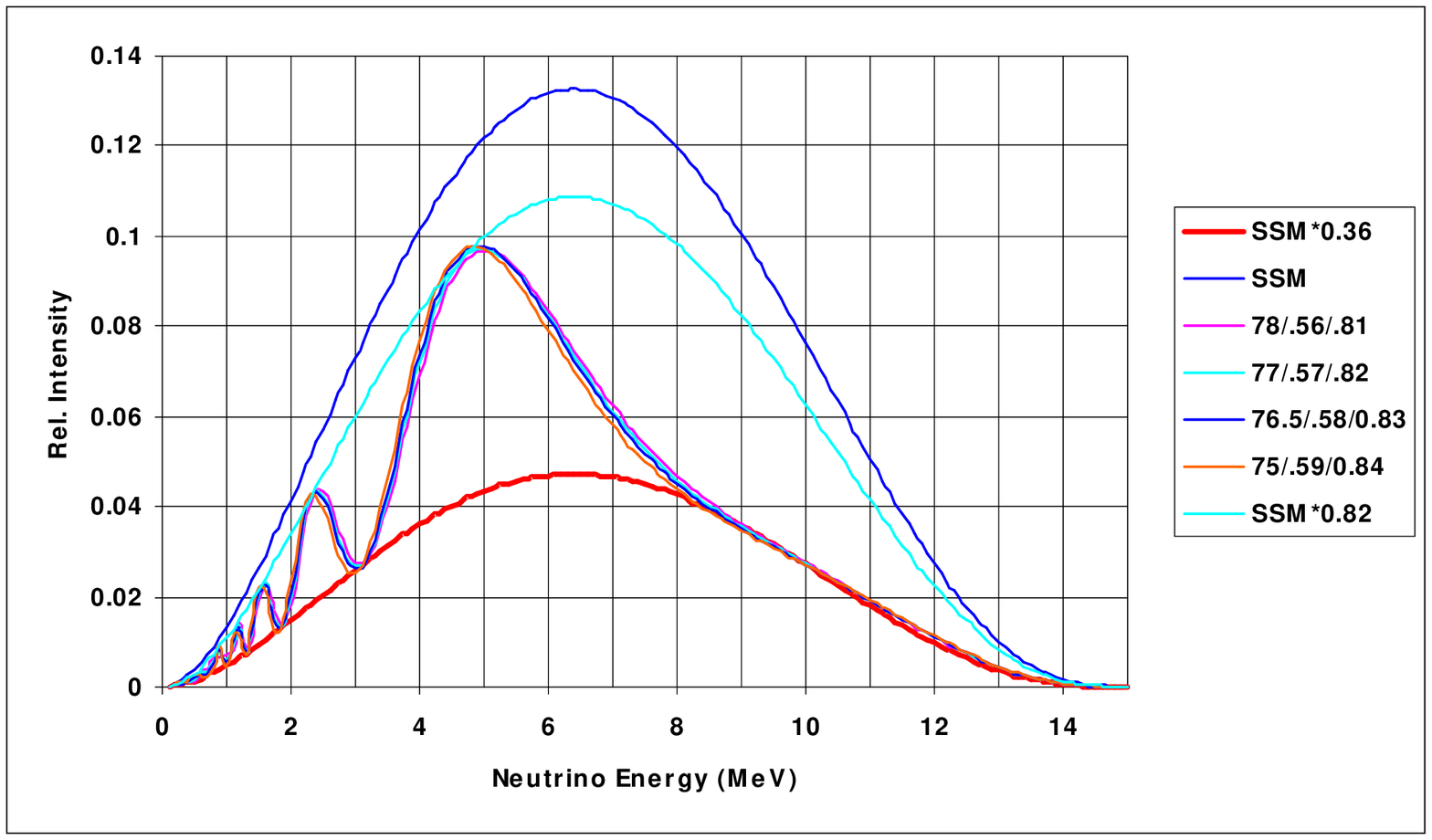,height=5.0in,width=6.5in}}
%\vspace{-2.5in}
\caption{$^8$B Solar neutrino spectra under different LWO scenarios
(the parameter combinations [$\Delta m^2 (\mu$eV$^2$)/sin$^2 2 \theta / \phi$(B)
(fraction of $\phi$(SSM)] are listed on the right).
The spectra are engineered to produce a ``shape invariant'' high energy part.
The central curve (red) is the LMA-MSW $\equiv 0.365 \phi$ (B:SSM). }
\end{figure}

{The shapes change extremely sensitively with the three }{\(\nu\)}$_{\mathrm{{e}}}${ parameters that tune up a viable solution systematically:
The }{\(\Delta\)}{m}$^{\mathrm{{2}}}${ arranges the flavor survival P}$_{\mathrm{{ee}}}${ spectrum such that a roughly horizontal part energywise,
is positioned in the $>$5 MeV regime; the }{\(\phi\)}{(B)parameter scales it into the LMA shape window; the sin}$^{2} 2 \theta${ that strongly affects the curvature, flattens it into a nearly undistorted shape.
The final angles that expose the new model are in general,
smaller than previously examined;
e.g., solutions with the same }{$\Delta$}{m}$^{2}${=76}{$\mu$}{eV}$^{2}${ but sin}$^{2}2 \theta \sim${0.8,
beyond the present range $\leq$ 0.58, failed earlier \cite{suzuki} and in this work.  }

{Fig. 1 shows that the needed shape invariance can be achieved with precision above $\sim$8 MeV below which the LWO shapes fan out deviating from the LMA-MSW shape.
We aim at neutrino reaction signals such as ES and CC$>$ 5 MeV using these input shapes.
Thus, deviations in Fig 1 between 5 and 8 MeV appear problematic.
In both ES-SK and CC-SNO that apply the Cerenkov technique, the experimental signals of the electron kinetic energy T range above 5 MeV\@.
In the quest for LWO }{\(\nu\)}$_{\mathrm{e}}${ shapes that generate ES/CC shapes that fit the data, the spectral smearing due to the inherent reaction as well as the detector energy resolution play a key role.
In the end, for both CC and ES, the overall spectral smearing is such that the dominant contribution to the ES and CC signal comes from E(}$\nu_\mathrm{e})>8${ MeV where the spectral match is close.
The possible deviations occur mainly near the 5 MeV signal threshold where background is strong in both SK and SNO.}

{ This effect can be understood in the case of the ES shape since the ES signal at a given electron energy is derived from the whole neutrino spectrum above that energy,
i.e. the reaction smearing effect is complete\cite{bahcall3}.
Even a large deviation e.g., at 5--6 MeV is damped down considerably in the ES spectrum.
In the case of the CC spectrum T$>$5 MeV which implies E(}{\(\nu\)}$_{\mathrm{{e}}}${)$>$6.44 MeV, the deviations look more worrisome since in this case one normally expects 1-to-1 correspondence of E(}$\nu_{\mathrm{e}}${) and T, thus no reaction smearing.
In reality, there is considerable reaction smearing even for the CC signal since in the reaction }{\(\nu\)}$_{\mathrm{e}}${ + d $\rightarrow 2$p + e, the proton recoils cannot be neglected\cite{bahcall4}.
The effect removes the E(}{\(\nu\)}$_{\mathrm{{e}}}${)-T correspondence significantly (although not to the extent in ES).
The final effect is shown in Fig. 2 where the light blue curve for no recoil corrections peaks at 8 MeV\@.
The recoil correction results in a large downshift to $<$7 MeV\@.
The detector resolution smearing produces a further downshift to $\sim$6.5 MeV\@.
Thus, also the CC signal shape is determined largely by E(}$\nu_{\mathrm{e}}${)}{$>$8 MeV\@.}

\begin{figure}[ht]
\vspace*{-1.5in}
\centerline{\epsfig{file=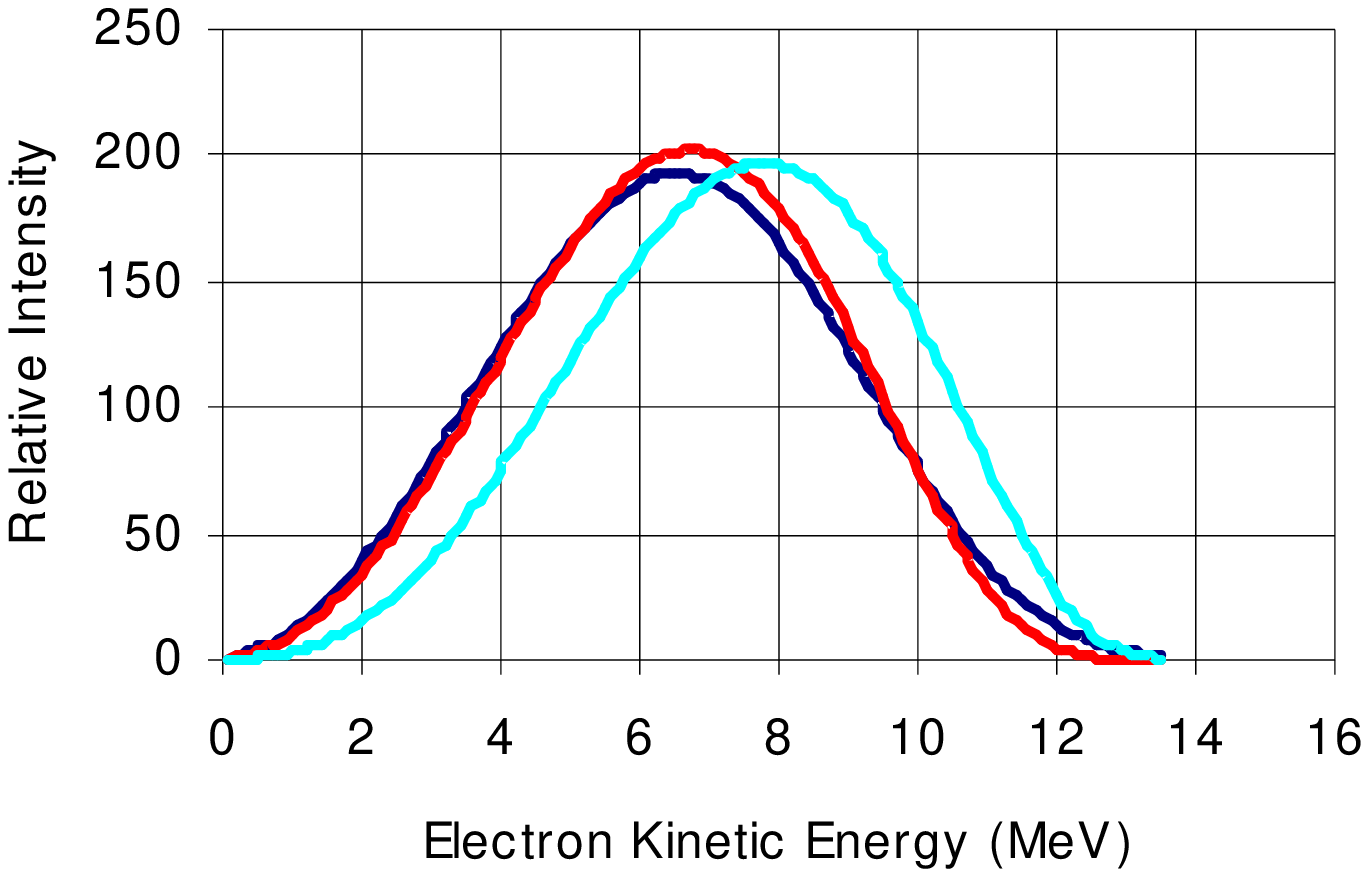,width=6.5in}}
\caption{CC spectral shapes from $\nu_\mathrm{e}$+d $\rightarrow$ 2p+e in SNO with 1) no recoil or detector resolution (light blue); 2) recoil only (red) and 3) recoil and resolution corrections (final experimental shape; dark blue).}
\end{figure}

{The plan of search for a new }{\(\nu\)}$_{\mathrm{{e }}}${ model was as follows.
In the first step, the viable basic parameters of the LWO scenarios were fixed by matching predicted shapes and rates to the ES-SK result, by far the most the stringent test.
The ES spectral/rate match set the primary fix of the sin}$^{2} 2 \theta${ and }{$\phi$}{(B) values for each selected }{\(\Delta\)}{m}$^{\mathrm{{2}}}${ in the viable band.
The match to the SNO CC data followed with the same parameter set except for possibly very slight changes in the scaling }{\(\phi\)}{(B) only.
With a reasonable match to data on }{\(\nu\)}$_{\mathrm{{e}}}${ (}$^{\mathrm{{8}}}${B), the matches to low energy data from }$^{\mathrm{{37}}}${Cl and }$^{\mathrm{{ 71}}}${Ga were tested with the same final set of parameters reached above.
The upper limit of the }{\(\Delta\)}{m}$^{\mathrm{{2}}}${ band is $\sim 80$}{$\mu$}{eV}$^{2}${,
because, beyond this limit, the spectral deviations around 5 MeV become too large to accommodate a spectral match to the data.  At $\sim${}70 }{\(\mu\)}{eV}$^{2}${ and below, the required scaling factor }{$\phi$}{(B) dips $<$0.75, the 2}{\(\sigma\)}{ limit of SNO. }

\section{Electron Scattering}

{The procedure involved the calculation of the ES spectra for LMA and LWO scenarios normalized to that for the SSM spectrum with no conversion.  Thus, the ES signal for each scenario is:}

\begin{equation}
{\mathrm{R(ES)}=[<\mathrm{W}( \lambda;\mathrm{P}_\mathrm{ee})+\mathrm{Z}(\lambda ;1-\mathrm{P}_\mathrm{ee})>]/<\mathrm{W}(\lambda)> }
\end{equation}
{where  W and Z are the differential ES cross-sections  for e- and }{\(\mu\)}{/}{\(\tau\)}{ neutrinos respectively \cite{bahcall3}, averaged over the SSM }{\(\nu\)}$_{\mathrm{{e}}}${ spectrum P(}$\lambda${) \cite{bahcall3} modified by the e-flavor survival function P}$_{\mathrm{{ee}}}${ that, in general, depends on  E(}{\(\nu\)}$_{\mathrm{{e}}}${). The $<  >$ in (1) indicate convolution with the detector resolution in SK\cite{nakahata}.}

{ For the LMA case, P}$_{\mathrm{{ee }}}${ is a }\textit{{constant =0.365}}{, close to 0.349(22) (relative to the SSM flux of }$5.05 \times 10^6${ cm}$^{2}${s}$^{-1}${), the value found in CC-SNO.
Then, W(}$\lambda$;P$_\mathrm{ee}$)= W($\lambda$)$\times$0.365 and Z($\lambda$;1-P$_\mathrm{ee}$) =W($\lambda$)(1-P$_\mathrm{ee} ) 0.16 \approx$ W($\lambda) \times 0.635 \times 0.16$,
{because of the nearly constant value of Z/W$\approx$ 0.16 for solar }$\nu_{\mathrm{e}}${  energies   $>$3 MeV\@.
This leads to R(ES;lma) 0.466 $\approx$0.465(15), the SK result \cite{fukuda} relative to the SSM flux above (the reason for choosing P}$_{\mathrm{{ee}}}${ (LMA)=0.365).}

{In the case of the LWO spectra, the same procedure was used. Here P}$_{\mathrm{{ee}}}${ is energy dependent, given by the standard oscillation formula for solar baseline:}

\begin{equation}
{\mathrm{P}_\mathrm{ee}=1-\mathrm{sin}^{2}2\theta \: \mathrm{sin}^2\{0.1899[\Delta m^2(\mu\mathrm{eV}^2)/\mathrm{E}(\nu_\mathrm{e})\mathrm{(MeV)}]\} }
\end{equation}

{                                                               }

{For each value of }$\Delta${m}$^2$, using the convoluted spectra $<$W($\lambda$;P$_\mathrm{ee}) + $Z($\lambda$;(1-P$_\mathrm{ee})>${ and $<$W(}$\lambda >${, the ES signals were calculated as in (1).}

{This spectrum (energy dependent in principle), was matched to the data points of the ES-SK spectrum \cite{vagins} by varying the parameters }{\(\phi\)}{(ES) ($<$1) and
sin}$^{2} 2 \theta${ for each value of }{\(\Delta\)}{m}$^{\mathrm{{2}}}${ in  the range }{}{70-80 }{\(\mu\)}{eV}$^{\mathrm{{2}}}${.
The best match was selected interactively for a given value of }{\(\Delta\)}{m}$^{\mathrm{{2}}}${ from the }$\chi^{2}${}{of the fit of the R(ES) curve to the ES-SK data points.
The results are shown in Fig. 3.  }

\begin{figure}[ht]
\vspace*{-2.5in}
\centerline{\epsfig{file=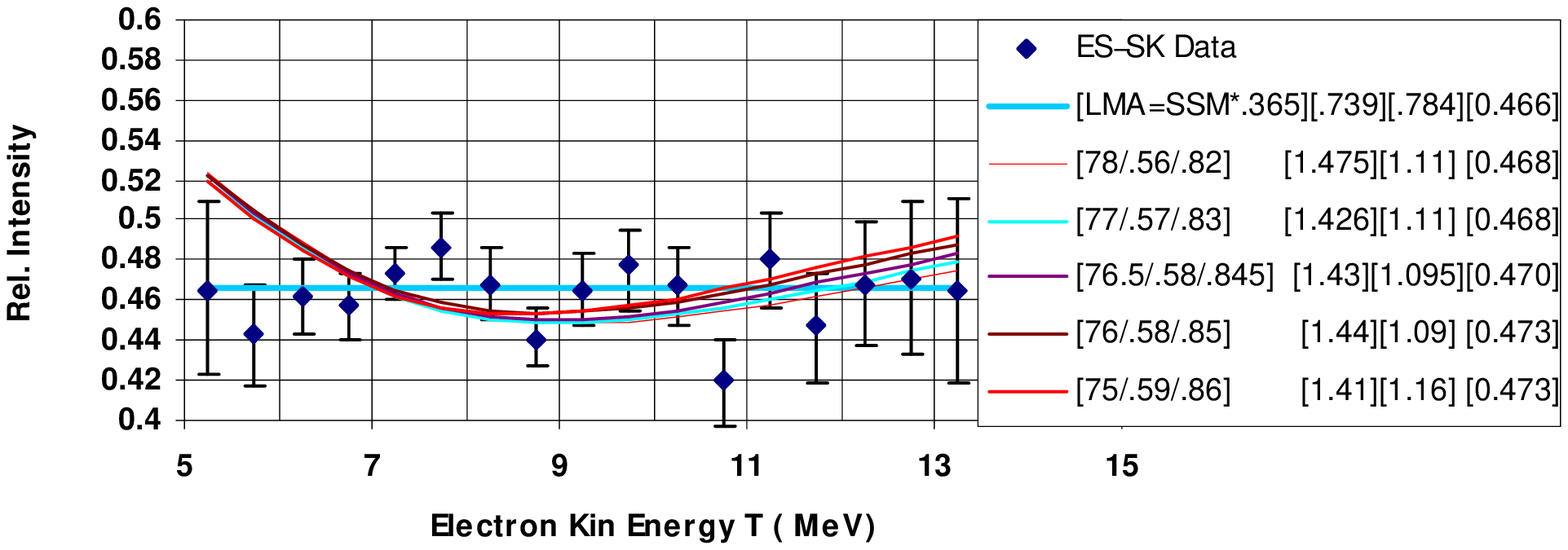,width=6.5in}}
\caption{R(ES) spectra of LWO matched to ES data points from SK. The legend box lists the parameter groups: 1) the best match $\nu_\mathrm{e}$ parameters
[$\Delta$m$^2$($\mu${eV}$^2$)/sin$^2 2 \theta / \phi$(ES)];
2) the [ $\chi^2$] values of the fit to the ES-SK data points in the energy range 5-14 MeV;
3) [$\chi^2]$ for the fit in the range 6.5--14 MeV;
and 4) the area of [R(ES)] as fraction of the SSM flux assuming no oscillations.
LMA = SSM*0.365 leads to the result R(ES-LMA)=R(ES-SK)=0.466.}
\end{figure}

{The box in Fig. 3 shows the values of the parameter combinations [}{\(\Delta\)}{m}$^{\mathrm{{2}}}${; sin}$^{2} 2 \theta${; }{\(\phi\)}{(ES)], ``the ``best match'' [}{\(\chi\)}$^{\mathrm{{2}}}${] (dof =16 for 17 SK data points in the range 5-14 MeV) and the  [R(ES) area] normalized to SSM for each curve including ES-lma. The ES(lwo) curves are seen to match the SK data, particularly those for }{\(\Delta\)}{m}$^{\mathrm{{2}}}${ = 75-78 }{\(\mu\)}{eV}$^{\mathrm{{2}}}${, even in the narrow slit allowed by the high statistical precision of the SK data. Indeed, except for the first two data points at the lowest energies in Fig. 3 that are the most susceptible to background,
the }$\chi^2${ values  (dof =14) for T= 6.5--14 MeV (also given in the box in Fig 3) for }{\(\Delta\)}{m}$^{\mathrm{{2}}}${ =75-78 }{\(\mu\)}{eV}$^{\mathrm{{2}}}${ jump to 1.05 to 1.15 with probabilities in the range of 35\% compared to the best fit 50\% value for }{\(\chi\)}$^{\mathrm{{2}}}${=1.
The rate matches are excellent (see Table 2 below). Our new model thus passes its most severe test.
A key reason is the significantly smaller mixing angles used here (compared to previous analyses) that flatten the ES spectra sufficiently to be scaled into the ES data window.
Another key effect is detector resolution as shown in Fig. 4.
The unconvoluted ES(lwo) curve (in blue) with a discouraging shape (even without the low energy points) reduces to the much more acceptable curve (in red) that includes the effect of the SK energy resolution.}

\begin{figure}[ht]
\vspace*{-2.5in}
\centerline{\epsfig{file=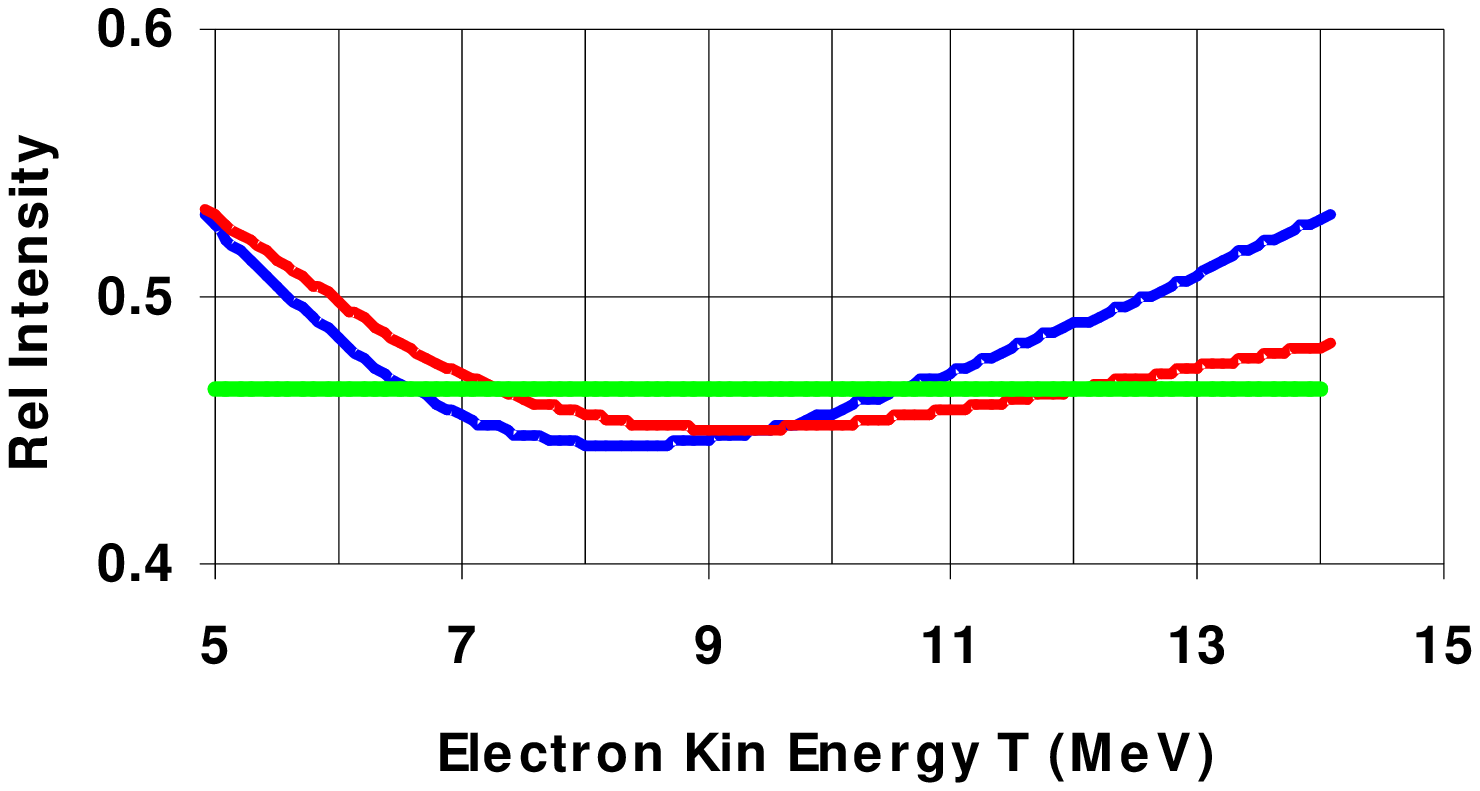,width=6.5in}}
\caption{Effect of detector resolution (SK) on ES(lwo) spectrum (blue). The convolution improves the match of the observed spectrum (red) to ES(lma) (green). }
\end{figure}

\section{CC Reaction on deuterons}

{The CC reaction }$\nu_{\mathrm{e}} +\mathrm{d} \rightarrow \mathrm{2p}+\mathrm{e}${ has been applied at SNO to measure directly the }$\nu_\mathrm{e}${ }$^{8}${B}{ spectrum.
The theoretical cross sections for this reaction including the recoils of the two protons in the final state have been extensively tabulated\cite{nakamura}.
The calculation of the CC spectra for LWO and LMA, input }{\(\nu\)}$_{\mathrm{{e}}}${ spectra must be summed over the contributions of }{\(\nu\)}$_{\mathrm{{e}}}${ from 0 to E(}{\(\nu\)}$_{\mathrm{{e}}}${)(max) = T+ Q (=1.44 MeV) for each signal electron kinetic energy T instead of just from the single }{\(\nu\)}$_{\mathrm{{e}}}${ energy T+Q.
The cross section C(}{$\lambda$}{; P}$_{\mathrm{{ee}}}${) is first averaged over the input solar }{\(\nu\)}$_{\mathrm{{e}}}${ spectrum P(}{\(\lambda\)}{) modulated by the P}$_{\mathrm{{ee}}}${ of the oscillation scenario.
P}$_{\mathrm{{ee}}}${(LMA) =0.365.
P}$_{\mathrm{{ee}}}${(LWO) is given by  (2).
For each value of }{\(\Delta\)}{m}$^{\mathrm{{2}}}${ the CC shapes were generated as:}

\begin{equation}
{\mathrm{R(CC)} = \phi (\mathrm{CC})<\mathrm{C}( \lambda; \mathrm{P}_\mathrm{ee}) >}
\end{equation}
{where P}$_{\mathrm{{ee}}}${ was calculated with the same mixing parameter fixed in the ES-SK spectral match.
The $<  >$ indicates, as before, convolution with the SNO detector resolution\cite{ahmad}. }{\(\phi\)}{(CC) is the scale factor applied to match the LWO-CC spectrum to LMA-CC spectrum measured at SNO.
The shape matches aimed at fits (by eye) within a $\pm$12\% band, the error band in the measured SNO-CC spectrum\cite{ahmad}.
Fig. 5 shows the LWO-CC matches to the LMA-CC spectrum (thick red line marked SSM*0.365).
The legend box shows in [ ] the signal areas under the curves in the window 5--13.5 MeV\@.}

\begin{figure}[ht]
\vspace*{-2.5in}
\centerline{\epsfig{file=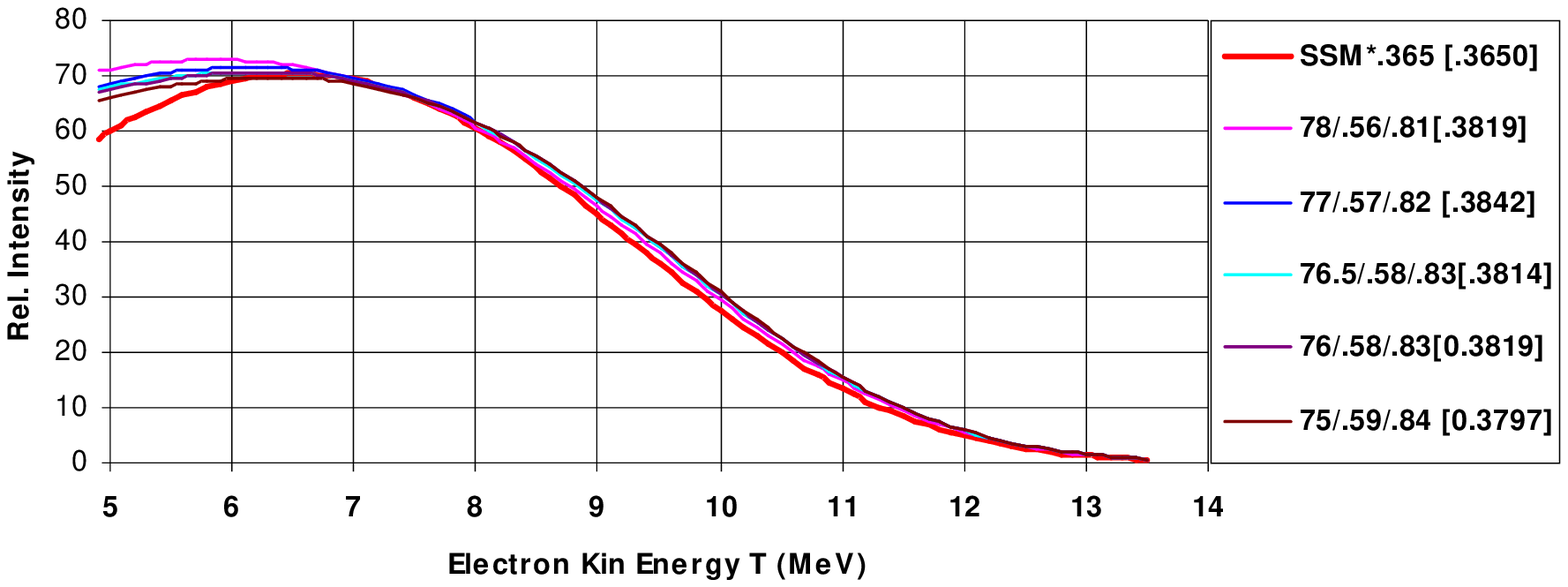,width=6.5in}}
\caption{CC spectra of $\nu_\mathrm{e}$+d$\rightarrow$2p+e for LWO models with parameter combinations listed in the box, compared with the prediction for LMA (=SSM*0.365; thick purple line). The SNO-CC measurement observes this curve with an error band $\sim \pm$12\% which completely encloses the LWO curves.  The areas normalized to SSM are shown in [ ].}
\end{figure}

{The scale parameter }{\(\phi\)}{(CC) is the true flux of  }{\(\nu\)}$_{\mathrm{{e}}}${(}$^{\mathrm{{8}}}${B).
The spectral matches in Fig. 5 are good with only a $\sim$10\% deviation even at the lowest energies that are most susceptible to errors from background and the disentanglement of the relatively broad NC }\textit{{\(\gamma\)}}\textit{{-ray}}{ }\textit{{shower}}{ signal that occurs at the effective energy of $\sim$5 MeV instead of at its full energy of 6.15 MeV\cite{chen}.
The rate matches are also good (see Table 2).
The best match LWO parameter group in the CC matches as well as overall, is [}$\Delta$m$^2$ ($\mu${eV}$^2$)/sin$^2 2 \theta / \phi${(CC)] = [77/0.57/0.82]. }

\section{Low Energy Signals from $^{37}$Cl and $^{71}$Ga}

{We now consider how the low energy parts of the new model, constructed so far only with  high energy }$^{\mathrm{{8}}}${B solar neutrinos in mind, match the signals from the  }$^{\mathrm{{37}}}${Cl and }$^{\mathrm{{71}}}${Ga detectors that respond not only to the low energy parts of the }$^{\mathrm{{8}}}${B spectrum (}$^{\mathrm{{37}}}${Cl Q=0.814 MeV) but beyond, to the lowest energies of  the pp }{\(\nu\)}$_{\mathrm{{e}}}${'s (}$^{\mathrm{{71}}}${Ga Q=0.235 MeV).}

{Interaction rates (in SNU=10}$^{\mathrm{{-36}}}${ /target nucleus/s) were calculated using well-known target data and standard }$\nu_{\mathrm{e}}${ spectra from every solar source \cite{bahcall5} modulated by the LWO effect using (2). Table 1 lists these results for }$^{\mathrm{{37}}}${Cl and }$^{\mathrm{{71}}}${Ga for the SSM as well as LWO with the same parameters fixed by the ES and CC-matches above. The values for the SSM agree with published values\cite{bahcall1}, checking the overall source data and the calculation methods.  }

{Table 1 shows the signal rates in the lines marked ``Total SNU'' in 4 different }\textit{{solar astrophysical }}{scenarios}.
{These rates}\textit{{ }}{compare with the measured SNU for Cl and Ga (last line in each section).
The favored LWO groups are highlighted.
The first result is the line marked SSM, for the standard sun.
The LWO results for Ga range between $\sim$80--82 SNU that deviate by $\sim 2\sigma$}{ from the measured value of 70.8(44) SNU\cite{sagegallex}.
The match of  LWO rates with the Cl data,  2.56(23) SNU\cite{cleveland}, is worse, deviating systematically from it by} $> 3 \sigma${ for every LWO parameter choice.
A similar problem exists }\textit{{also for the LMA-Cl match }}{(see Table 2 below).
The fact that the mismatch in Cl is much worse than in Ga indicates that problem lies in the spectral region between 1--5 MeV\@.
Neutrino spectral engineering cannot solve this problem.}

{Astrophysical spectral adjustments, on the other hand, could be resorted to, to a limited extent by changes in the solar fluxes.
The only negotiable fluxes in the framework of the SSM are the }$^{\mathrm{{8}}}${B flux (predicted with a large error $\sim$20\%) and the CNO fluxes ($\sim$100\%)\cite{bahcall1}.  Two astrophysically sound remedies touching these solar sources are conceivable.}

{The LWO scenarios above are based inherently on the assumption }$\phi${(B)}$< \phi${(B:SSM).
This idea must be astrophysically based, not stand as an ad hoc construct as tacitly viewed so far.
The simplest way to reduce the }$^{\mathrm{{8}}}${B flux is to reduce the central temperature T of the sun using astrophysically sound scaling laws: }{$\phi$}{(B)$\propto$T}$^{20}${ and }{$\phi$}{(Be)$\propto$T}$^{10}${\cite{fiorentini}.
Using the matched values of }{\(\phi\)}{(CC-B) above ($\sim$71--80\%), the reduction in }{\(\phi\)}{(Be) can be calculated.
This correction is applied in the lines marked SSM(T) in Table 1.
The correction helps to a small extent.
}

{The second remedy concerns the CNO flux for which there is little astrophysical guidance.
Global fits of all neutrino scenarios proposed so far are eminently compatible with 0--5\% CNO luminosity in the sun\cite{bahcall6}.
We therefore suggest that the CNO flux is seriously reduced, viz. =0.
The lines marked (No CNO) reflect this assumption.
This correction helps substantially in the LWO-Cl match.
Finally both astrophysical modifications result in the rates marked SSM [No CNO(T)].
The Cl rates now match the LWO predictions within 1.65}{\(\sigma\)}{ (similarly, no-CNO helps }\textit{{also the LMA-Cl}}{ match to move within 1.5}{\(\sigma\)}{, see Table 2) The LWO Ga match also improves to a much closer $\sim$0.3}{\(\sigma\)}{.
We therefore accept and  }\textit{{require }}{these astrophysical modifications; hence the designation ``astroparticle'' to our model. }

{Table 1 indicates the reason why }{\(\Delta\)}{m}$^{\mathrm{{2}}}${ values below 75}{\(\mu\)}{eV}$^{\mathrm{{2}}}${ become rapidly untenable.
The LWO oscillations become progressively faster as E(}{\(\nu\)}$_{\mathrm{{e}}}${) decreases (see Fig. 1).
At the energy of the }$^{\mathrm{{7}}}${Be line (0.862 MeV), the oscillation period changes rapidly with }{\(\Delta\)}{m}$^{2}$.
{Below 75 }{\(\mu\)}{eV}$^{\mathrm{{2}}}${ the Be line strobes into a resonance, thus P}$_{\mathrm{{ee}}}${ jumps to $\sim$1.
The Be contribution to the Cl (as well as the Ga) rate jumps, causing both to fall out of match.
Even though the LWO-}$^{\mathrm{{8}}}${B matches are good for }$85 > \Delta${m}$^2 > 70 \mu${eV}$^2${ the Be factor limits the model at }{\(\Delta\)}{m}$^{\mathrm{{2}}}${ = 75 }{\(\mu\)}{eV}$^{\mathrm{{2}}}$ as long as $\phi$($^8$B) $\ge$
0.8 $\phi$($^8$B:SSM).

\section{Model \& Data}

{The discussion above has examined in detail the compatibility of the predictions of the new LWO-based astroparticle model of solar neutrinos to every experimental result available to date with sensitivities that span the entire solar neutrino spectrum.
In Table 2 we summarize the match of model to data quantitatively and individually.
The ``astroparticle LWO model'' is competitive with the (astroparticle) LMA in the consistency with the global data.
The major}\textit{{ }}{difference is a 1.4}{\(\sigma\)}{ mismatch with the SNO }\textit{{absolute NC rate}},
{a main reason why the LWO scenario was disfavored in global analyses
such as \cite{fogli} even though it was found competitive
in the absence of the NC result.}

\section{Experimental tests}

{ For the LMA-MSW, }\textit{{only}}{ two types of experimental data can provide new support.
The first is a substantial improvement in the precision of the absolute NC signal rate.
We note that the fundamental difference }\textit{{vis a vis}}{ the LWO model is a mere 20\% lower }$^{\mathrm{{8}}}${B }{\(\nu\)}$_{\mathrm{{e}}}${ flux.
The NC rate in SNO is now known to $\pm$13\% 1}{$\sigma$\cite{ahmad}.
This precision needs to improve markedly to exclude the possibility of a 20\% lower flux by at least 3}{\(\sigma\)}{.
The second possibility is the only predicted effect specific to the LMA viz.,
a gradual rise of the CC signal from 0.365 SSM above 5 MeV to 0.6 SSM below 1 MeV in the region of the }$^{7}${Be and pp neutrinos\cite{bahcall7}.
This regime, endemic to high background is difficult to access experimentally and the LMA vs. LWO test requires sharp differentiation of somewhat comparable signals at 0.6 SSM and 0.45 SSM (see last lines in Table 1) respectively.
The first opportunity for this test will arise when Borexino\cite{borexino},
designed for very low background at low energies, starts operation in 2004.
We now turn to experiments aimed specifically at exposing the LWO basis of the new model.}

\section{ES Signals at KamLAND, BOREXINO}

{The basic premise of the present model is spectral shape invariance for E(}{\(\nu\)}$_\mathrm{e}) > 5${ MeV under the action of LWO.
However, as seen in Fig. 1 this action is fundamentally accompanied by }\textit{{strong shape changes}}{ }\textit{{immediately below}}{ }\textit{{5 MeV\@.}}
{ Experimental tests accessible to the signal window $<$ 5 MeV, hitherto unexplored via direct solar neutrino detection, hold the key to a compelling demonstration of the present model since the LWO spectral predictions in this window are dramatically different from the relatively weak  shape effects in the LMA case.
This window is closed to the present large-scale Cerenkov detectors SK and SNO.
It is however accessible to a liquid scintillator (LS) based detector on the scale of at least 1 kton target mass (similar to SNO though much smaller
than SK with 22 kton  fiducial mass).
KamLAND is such a detector that is already operational\cite{fukuda}.
Borexino is another, that will soon be operational.
Both are designed for low background neutrino spectroscopy at low energies, a key factor in considering signals below 5 MeV\@.
We first consider KamLAND because interesting data can be expected from this device in the very near future.}

\begin{figure}[ht]
\vspace*{-2.5in}
\centerline{\epsfig{file=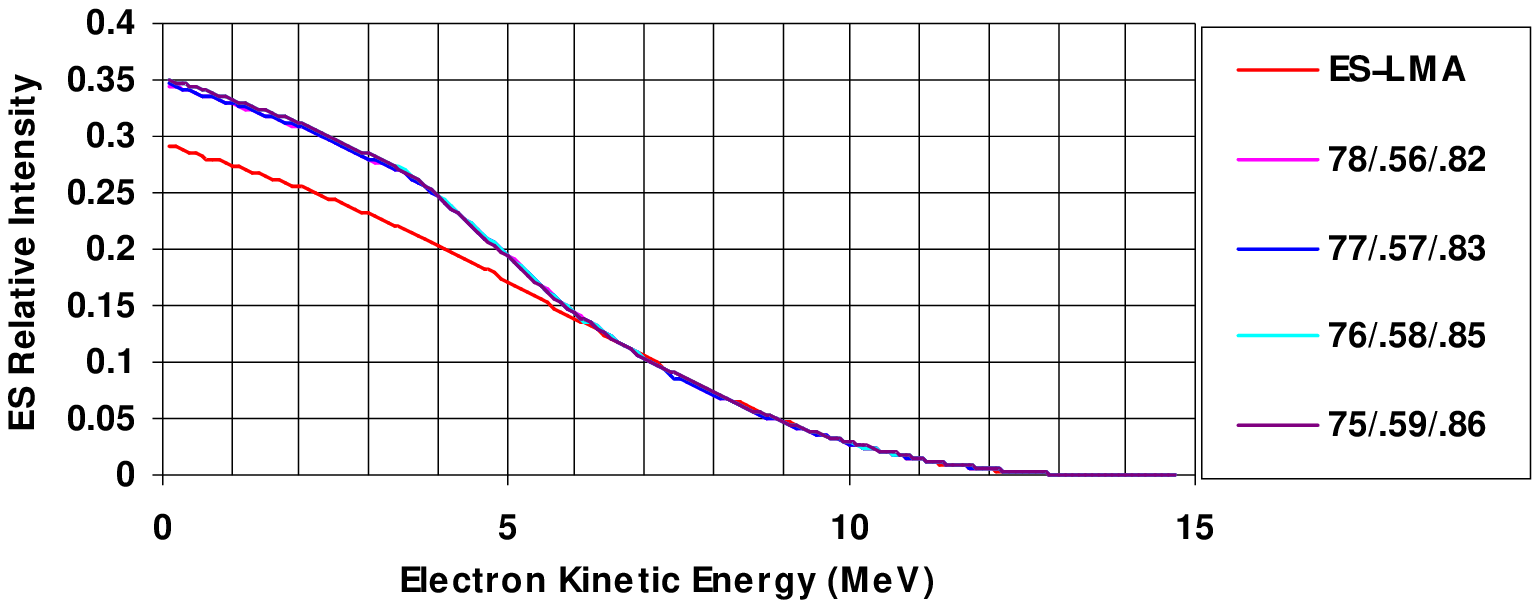,width=6.5in}}
\caption{ES spectra predicted for  LMA and LWO in the KamLAND liquid scintillation detector. The curves are unconvoluted with energy resolution  that is not expected to affect the shapes significantly in KamLAND\@. }
\end{figure}

{ Let us first consider ES signals below 5 MeV\@.
Fig. 6 shows the ES spectra expected in a 1 kton LS detector.
These are theoretical curves unconvoluted with detector resolution since the superior energy resolution in a LS device results in little smearing of the spectral shapes.
Fig. 6 shows that while the ES spectra for LMA and LWO match identically above 6 MeV, they diverge significantly below this critical energy.
While it is obviously desirable to measure the full spectrum and determine the shape characteristics below 6  MeV (normalized  to the ``flat'' part $>$6 MeV), a reliable preliminary indication of the effect can be obtained just from the relative rates in two broad windows below and above 6 MeV\@.
Table 3 lists the signal rates/2y live time for the two models in the windows 3--6 and 6--16 MeV\@.
In the LWO scenario (with practically no differences between the viable solutions), the excess of counts at 3--6 MeV is measurably higher than in LMA, 484(39) vs. 370(38), a $\sim 3 \sigma$}{ effect.
For the lower threshold at 2.5 MeV background permitting, the LWO effect can possibly be detected at the $\sim 4 \sigma$}{ level.
KamLAND has been in operation now for nearly two years but it is not clear if data on signals in these windows are available.
Information is therefore eagerly awaited.}

{In KamLAND the background characteristics in the 2--5 MeV regime, the key factor, is not known.
The relatively shallow depth of this detector could pose a high cosmogenic background interference.
In Borexino, with a larger overburden, this problem is less acute.
The disadvantage is that in the present design, the largest mass available in Borexino is only 300 tons\cite{borexino}.
It is envisioned that the target mass can be upgraded to approach 1 kton in the future, when a definitive result on LWO can be attempted.}

\section{$^7$\textbf{Li CC Spectroscopy}}

{The ES spectra in Fig. 6 are interesting and informative especially since they may be observable already in KamLAND\@.
However, the spectral effects are considerably less prominent than could be expected from Fig. 1 because of the basic spectral smearing effect of the ES reaction (although the less serious detector smearing is avoided).
One can thus expect stronger effects in CC transitions with bound nuclei in the final state that ensure that the signal electron energy is uniquely connected to the }$\nu_{\mathrm{e}}${ energy (unlike in the }{\(\nu\)}$_{\mathrm{{e}}}${+d reaction).
Other considerations in the choice of the reaction are: 1) a high rate and/or a large target mass since the }$^{\mathrm{{8}}}${B }{\(\nu\)}$_{\mathrm{{e}}}${ flux is small and 2) a low threshold.
The latter is important since an energy threshold $>$3 MeV is desirable in order to avoid background.
An ideal choice is the case of }$^{\mathrm{{7}}}${Li, long known as a candidate for direct solar neutrino spectroscopy \cite{reines_woods} but not so far applied in practice.}

\begin{figure}[ht]
\vspace*{-2.5in}
\centerline{\epsfig{file=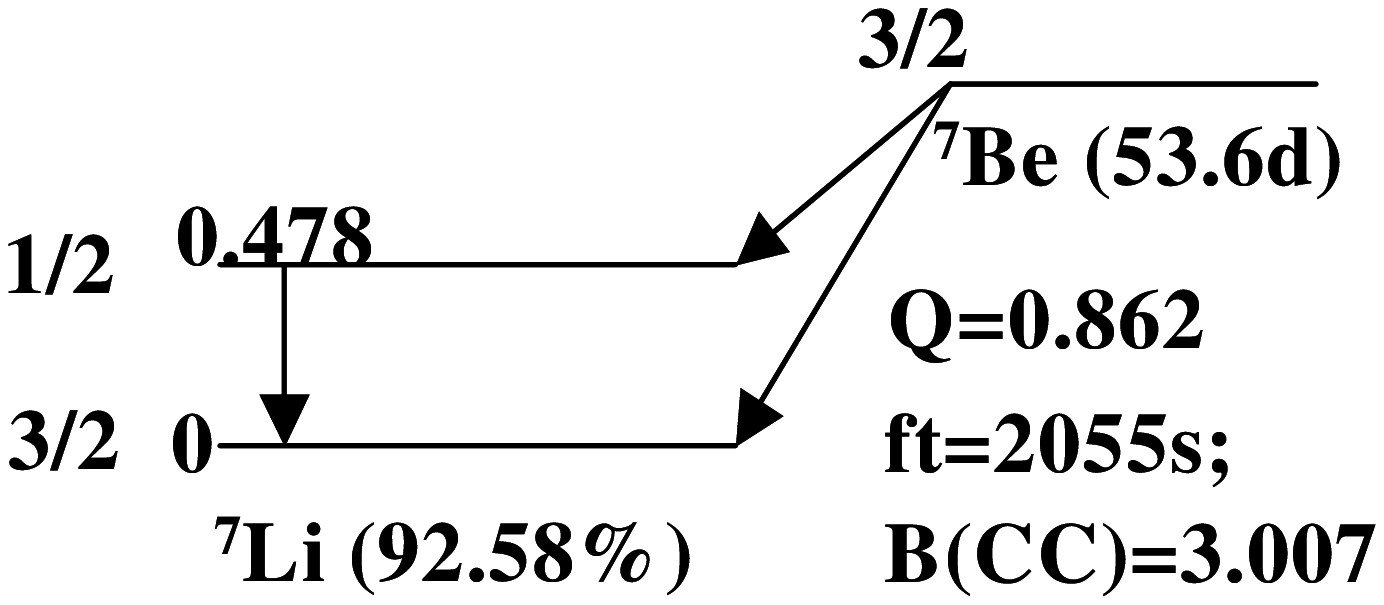,width=6.5in}}
\caption{Level scheme of the $^7$Li-$^7$Be Inverse $\beta$ system}
\end{figure}

{Fig. 7 shows the nuclear data relevant to the inverse }{$\beta$(IB)}{ reaction }{\(\nu\)}$_{\mathrm{{e}}}${+}$^7${Li}$\rightarrow ^7${Be + e.
The IB threshold is only 0.862 MeV so that dominant part of the }$^{\mathrm{{8}}}${B }{\(\nu\)}$_{\mathrm{{e}}}${    spectrum can be observed above a $\sim 3$ MeV signal threshold.
The IB transition is a superallowed transition with a precisely calculable rate, one of the highest known in beta decay.
As a light nucleus with a high natural abundance, a relatively small mass of 30--100 tons (compared to 1 kton) may suffice for a decisive result.
Technical details are beyond the scope of this paper, but a suitable technology for the Li-CC experiment appears possible.}

\begin{figure}[ht]
\vspace*{-2.5in}
\centerline{\epsfig{file=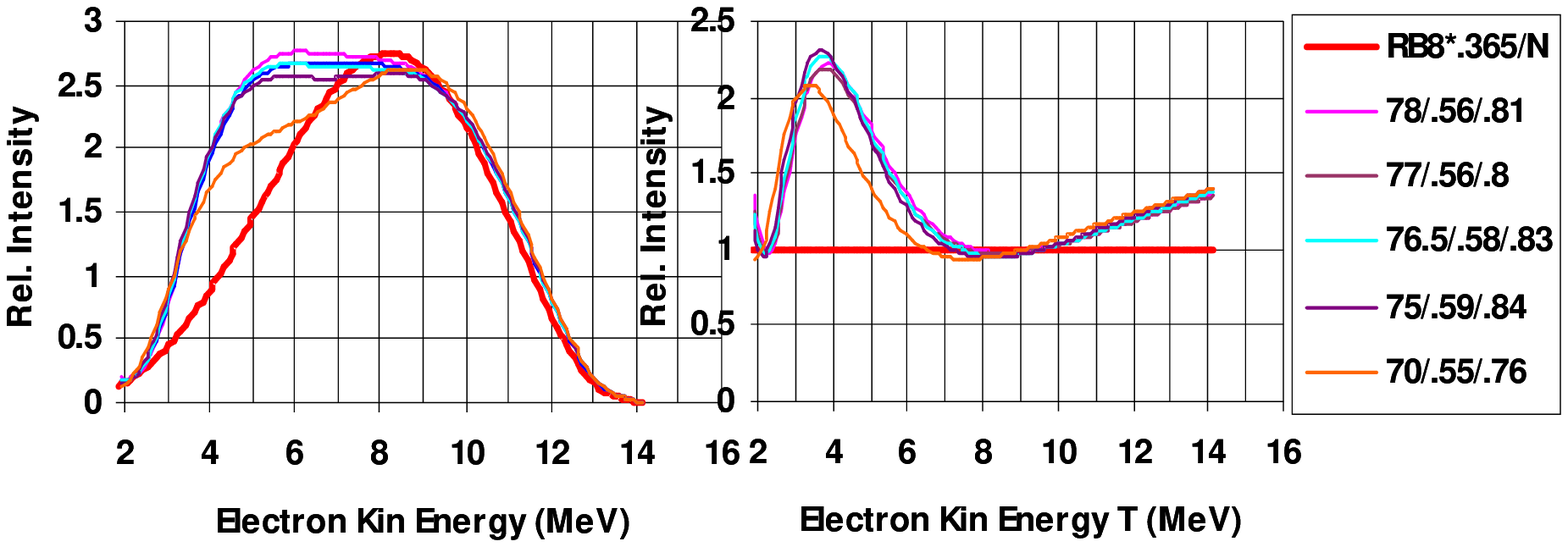,width=6.5in}}
\caption{Theoretical CC spectral shapes from the reaction $\nu_\mathrm{e}+^7$Li$\rightarrow ^7$Be + e (Q=0.862 MeV) for different LWO parameters and LMA=SSM*0.365 (thick red).
The panel on the left shows the inverse beta reaction spectra and
on the right, the spectra normalized to that of SSM*0.365. }
\end{figure}

{Fig. 8 (left) shows CC spectra observable in the Li IB reaction.
Almost the entire spectrum occurs above 3.5 MeV, comfortably above the background threshold.
With the thick red line for SSM*0.365=LMA as the reference, the spectra directly display the essential action of Fig. 1, --shape invariance above 8 MeV and dramatic modulation at lower energies.
The prominence of the modulations is illustrated in Fig. 8 (right) where the LWO spectra are normalized to LMA which appears as a flat line.
The spectral deviations are strong and compelling.
Table 4 lists the signal rates in two broad windows above and below 7 MeV\@.
The yearly rates are high.
In the low window, the LWO rates are $\sim$50\% higher than the LMA, whereas in the high window they are same (within $\sim$3\%).
It would thus appear that the essential proof of the LWO basis of the present model could be obtained in one year's live time possibly with a target mass as small as 30 tons of Li.}

{The LWO modulations become even stronger at very low energies in regime of the pp neutrinos.
Even in the relatively small window of a few hundred keV several periods of LWO modulations occur.
These modulations }\textit{{may }}{be observable by a suitable CC transition with a sufficiently low threshold that can be implemented by a technology that offers sharp energy resolution.
The modulations at these energies are more sensitive to the neutrino parameters than at the high energy windows considered above.
Thus it may be possible to determine them more precisely than the narrow LWO specifications already inherent in the present model.
An ideal choice for this application is the CC transition in }$^{\mathrm{{115}}}${In \cite{raghavan2}.
A practical real time pp neutrino detector based on indium is under development in the LENS project \cite{LENS}.
The application of LENS to the spectroscopy of the present neutrino model will be discussed elsewhere.}

\section{Conclusions}

{Is the LMA-MSW model the last word in solar neutrinos? It may well be.
The present work raises a serious possibility that it need not }\textit{{necessarily }}{be.
We have made the }\textit{{prima facie}}{ case that a model based on LWO, a very different physics scenario, offers a credible alternative.
Only a 20\% difference in a measured absolute signal rate stands between them.
The serious pause against the new model is, clearly, the possible violation of CPT invariance that may hold rigidly also in the neutrino sector in spite of recent suggestions to the contrary.
It is thus a classic case of decision by a unique physical result vs. discovery by consensus with a mass of different data.
Experiments with the potential for compelling spectral impact are proposed in this work and may already be observable.
Only they can tell which model is the prevalent one.}

{The impact of this decision goes well beyond the already rich field of solar neutrinos.
This science has led us deep into the sun's interior and beyond the Standard Model of particle physics.
Is it poised to take us to yet more unknown territory opened by the breakdown of one of the most cherished symmetry principles in physics?}

\section{Acknowledgments}

{I wish to express my sincere thanks to Peter Kurcynski who initiated me into the mysteries of MS-EXCEL computing used extensively in this work and for helpul discussions.
I am grateful to K. Kubodera and S.Nakamura for generously supplying their tabular data in a format ready for use in this work.
I thank John Bahcall, Michel Cribier, Eligio Lisi, Sandip Pakvasa and  Carlos Penya-Garay for their comments on the draft MS.}

\newpage
\begin{table}[ht]
\caption{LWO Predictions for the $^{37}$Cl and $^{71}$Ga solar neutrino detectors. The viable groups are highlighted and the best overall LWO choice is highlighted deeper.}
\vspace{-1in}
\centerline{\epsfig{file=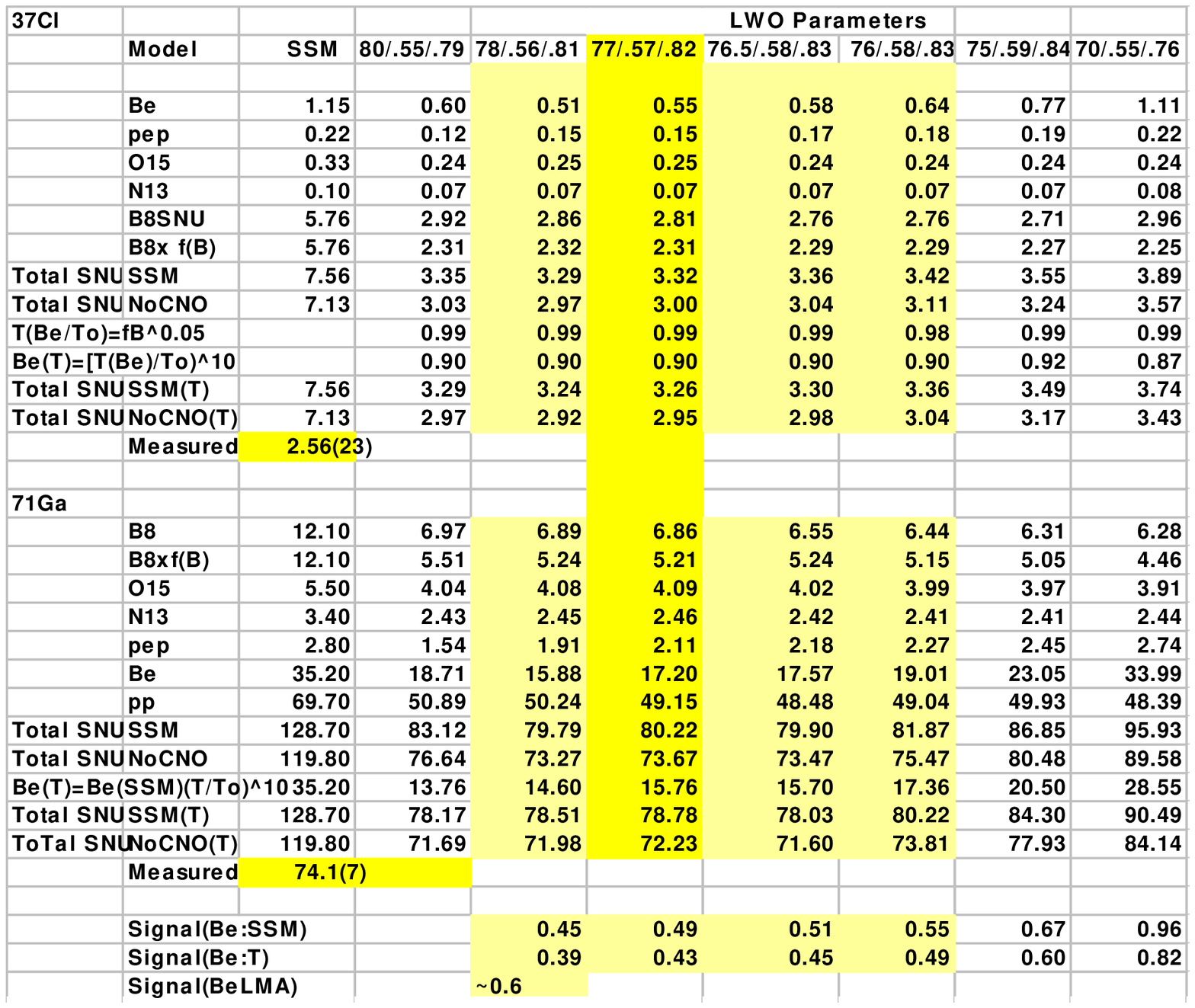,height=7in}}
\end{table}

\begin{table}[ht]
\caption{Comparison of LWO and LMA vs. Experimental Data}
\vspace{-2in}
\centerline{\epsfig{file=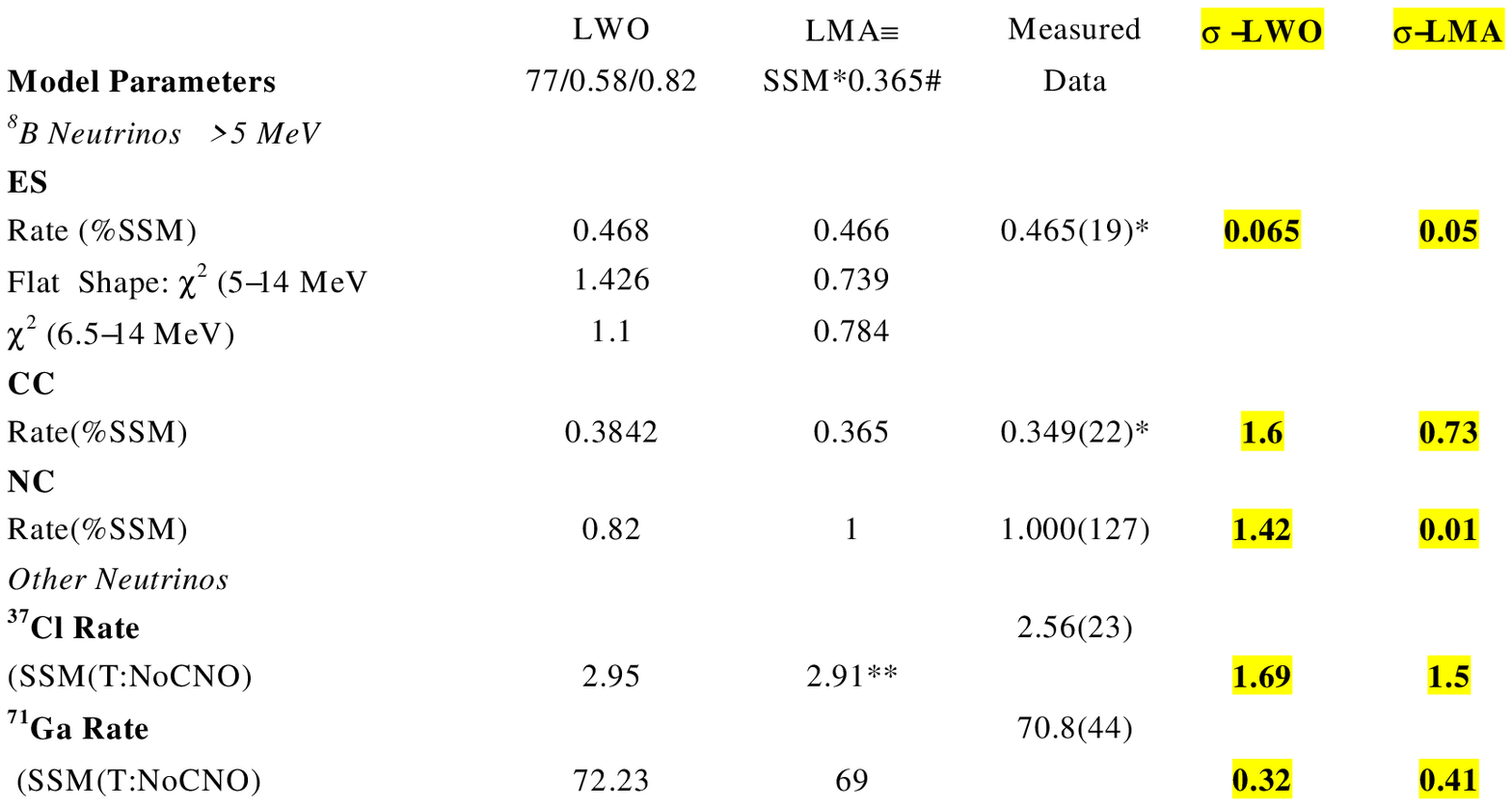,width=7.0in}}
\vspace{-2.0in}
\#  At E($\nu_e$) $>$ 5 MeV there is practically no difference between true LMA and SSM$\times$0.365\\*
*  Relative to BPSSM $\phi$($^8$B) = 5.05$\times 10^6/$cm$^2$s \\*
\$  True LMA-MSW \\*
** No CNO only
\end{table}

\begin{table}
\caption{ES Signal rates expected in KamLAND (1 kton liquid scintillator)/2 years live time}
\begin{tabular}{cccccc}
\hline
{Model}&{R(6-15 MeV)}&{R(3-6 MeV)}&{$\Delta$}{R}&{R(2.5-6 MeV)}&{$\Delta$}{R} \\
\hline
{LMA*}&{538}&{908}&{+370$\pm$38}&{1056}&{+518$\pm$40} \\
{LWO}&{536}&{1020}&{484$\pm$39}&{1220}&{+684$\pm$41} \\
\hline
\end{tabular}
% \vspace{0.3in}
*  SSM$\times$0.365, a sufficiently close approximation in the energy windows considered here.
\end{table}

\begin{table}
\caption{Solar neutrino signal rates/year in low and high energy windows for a 100t Li CC detector}
\begin{tabular}{ccccccc}
\hline
Window (MeV)&SSM*0.365&7.8/.56/.81&7.7/.56/.80&7.65/.58/.83&7.5/.6/.8&7.0/.54/.71 \\
\hline
3.5--7&1592&2480&2407&2434&2371&2027 \\
7--14&3213&3324&3291&3281&3270&3310 \\
\hline
\end{tabular}
\end{table}


\begin{thebibliography}{99}
\bibitem{nakahata}
{ The detector resolution function at SK was kindly supplied by  M.Nakahata.}
\bibitem{sagegallex}
{The cited value is the combined result of SAGE and GALLEX/ GNO as given in ref. 6.}
\end{thebibliography}
\end{document}